\documentclass[aps,pra,showpacs,twocolumn,superscriptaddress]{revtex4-1}
\usepackage{graphicx}
\usepackage{amsmath}
\usepackage{amssymb}
\usepackage{amsfonts}
\usepackage{bbm}
\usepackage{bm}
\usepackage{braket}
\usepackage{verbatim}
\usepackage{cancel}
\usepackage{tikz}
\usepackage{wasysym}
\usepackage[bookmarks=false,colorlinks=true,urlcolor=blue,citecolor=blue,linkcolor=blue]{hyperref}
\usepackage[normalem]{ulem}
\allowdisplaybreaks

\tikzstyle{tensor}=[rectangle,draw=blue!50,fill=blue!20,thick]

\begin{document}

\title{Nonergodic Quantum Dynamics from Deformations of Classical Cellular Automata}

\author{Thomas Iadecola}
\affiliation{Department of Physics and Astronomy, Iowa State University, Ames, Iowa 50011, USA}

\author{Sagar Vijay}
\affiliation{Department of Physics, Harvard University, Cambridge, Massachusetts 02138, USA}
\affiliation{Department of Physics, University of California, Santa Barbara, California 93106, USA}

\date{\today}

\begin{abstract}
Classical reversible cellular automata (CAs), which describe the discrete-time dynamics of classical degrees of freedom in a finite state-space, can exhibit exact, nonthermal quantum eigenstates despite being classically chaotic.
We show that every classical CA defines a family of generically non-integrable, periodically-driven (Floquet) quantum dynamics with exact, nonthermal eigenstates.
These Floquet dynamics are nonergodic in the sense that certain product states on a periodic classical orbit fail to thermalize, while generic initial states thermalize as expected in a quantum chaotic system.  We demonstrate that some signatures of these effects can be probed in quantum simulators based on Rydberg atoms in the blockade regime.
These results establish classical CAs as parent models for a class of quantum chaotic systems with rare nonthermal eigenstates.
\end{abstract}

\maketitle

Quantum many-body systems can support a rich variety of dynamical regimes.  In interacting many-body systems obeying the eigenstate thermalization hypothesis (ETH), unitary dynamics from a generic initial state leads to long-time steady states that are locally indistinguishable from thermal states~\cite{Deutsch91,Srednicki94,Rigol08,D'Alessio16}. Thermalization can be arrested altogether in the presence of strong disorder~\cite{Basko06,Oganesyan07,Znidaric08,Pal10,Nandkishore15,Abanin19} or substantially slowed in the absence of disorder due to the existence of rare eigenstates that fail to exhibit thermal properties vis-\`a-vis the ETH~\cite{Biroli10,Shiraishi17,Mondaini18,Shiraishi18a,Turner18a,Schecter19,Sala19,Khemani19}.  While many examples of such systems are known to exist, finding general criteria that determine whether a given class of many-body systems exhibits nonthermalizing dynamics remains an outstanding challenge.

Nonthermal quantum states also naturally arise as  eigenstates of classical, reversible cellular automata (CAs), which describe the discrete-time evolution of classical degrees of freedom in a finite state-space.    A reversible CA can equivalently be thought of as a periodically-driven (or ``Floquet") quantum system evolving under repeated application of a unitary Floquet operator $U_{\rm aut}$ that generates no entanglement in a certain basis of product states, which we refer to as the computational basis (CB)~\cite{Gopalakrishnan18,Shiraishi18,Iaconis19}.  In a finite system, every CB product state $\ket{\psi}$ thus has an exact recurrence under $U_{\rm aut}$ so that $U^\ell_{\rm aut}\ket{\psi}=\ket{\psi}$ for some $\ell$. Eigenstates of $U_{\rm aut}$ can be constructed by taking superpositions of the product states in such a closed ``orbit" of CB states. 

These unique properties of CAs can be used to make precise statements about eigenstate properties.  For example, the orbit length $\ell$ of an eigenstate of $U_{\rm aut}$ upper-bounds the entanglement entropy $S_A$ of any subsystem $A$ as $S_A\lesssim\ln\ell$~\cite{Gopalakrishnan18,Sarang_Automaton1}.  
In chaotic cellular automata defined on a 1D lattice with $L$ sites, typical orbits have $\ell\sim e^{c L}$~\cite{Takesue89}, so this bound is consistent with typical eigenstates having volume-law entanglement, similar to what is found in Floquet systems obeying the ETH~\cite{Kim14,D'Alessio14,Lazarides14,Ponte15}. However, even chaotic automata can have rare ``short" orbits for which $\ell$ is independent of system size; in the Supplemental Material (SM)~\cite{SM}, we show that all translationally-invariant automata have such short orbits. Eigenstates associated with these orbits must have area-law entanglement, and dynamics of initial states on these orbits exhibit classically nonergodic dynamics featuring finite recurrence times in the thermodynamic limit. 


In this work, we consider families of fully quantum-mechanical Floquet dynamics that generate, at a special point, the dynamics of a classical CA.  These quantum dynamics preserve certain nonthermal eigenstates of the classical CA while being otherwise quantum chaotic, revealing a connection between classical and quantum nonergodicity. 
We argue on general grounds that any CA with short orbits in one spatial dimension can be deformed in this manner, giving rise to ergodicity breaking in the sense of quantum many-body scars (QMBS)~\cite{Turner18a}. 
Since nonthermal eigenstates of a CA are simple to find numerically, 
our prescription provides a constructive approach to generating quantum-chaotic Floquet dynamics wherein certain states fail to thermalize.   

Our primary example of such dynamics is an integrable CA, which we show can be deformed into a nonintegrable Floquet model with exact area-law-entangled eigenstates descending from a short CA orbit. We demonstrate that quantum dynamics proximate to this CA can be realized using Rydberg-atom quantum simulators~\cite{Bernien17,Lienhard18,Guardado-Sanchez18}, where slow quantum dynamics and other signatures of the underlying CA can be observed under quantum quenches from different initial states. Our results complement recent studies of QMBS in a different nonequilibrium regime of the same physical system~\cite{Turner18a,Turner18b,Ho19,Choi18,Iadecola19,Michailidis19}, where slow dynamics arise by a different (incompletely understood) mechanism. Moreover, this work is distinct from recent examples of QMBS in Floquet systems~\cite{Mukherjee19,Haldar19,Sugiura19,Mizuta20,Mukherjee20} in that it provides a recipe for weak ergodicity breaking that can be applied in various physical contexts. 


\emph{\textbf{Nonergodic Quantum Dynamics and Classical Automata:}} Families of Floquet quantum dynamics that include a classical CA as a special case can exhibit nonergodic behavior. Consider the Floquet unitary operator $U_{\mathrm{aut}}$ that describes the evolution of a classical CA.  Any eigenstate of $U_{\rm aut}$ can be constructed by following the classical orbit of a CB product state $\ket{\psi}$ as:
\begin{align}
\label{eq:eigenstate}
\ket{\varepsilon_{\ell,p};\psi}=\frac{1}{\sqrt{\ell}} \sum^{\ell-1}_{k=0}e^{i\,\varepsilon_{\ell,p}k}\, U^k_{\rm aut}\ket{\psi},
\end{align}
where $\varepsilon_{\ell,p}=2\pi p/\ell$ ($p=0,\dots,\ell-1$) is the quasienergy of the state $\ket{\varepsilon_{\ell,p};\psi}$ associated with its eigenphase $e^{-i\varepsilon_{\ell,p}}$, and $\ell$ is the length of $\ket{\psi}$'s orbit under $U_{\rm aut}$.
A family of nonergodic quantum dynamics containing $U_{\rm aut}$
arises from the fact that there is a large family of nonintegrable deformations of 
$U_{\mathrm{aut}}$ that preserve certain short-orbit eigenstates as exact eigenstates. 
Since the short-orbit eigenstates have a constant entanglement entropy, they are guaranteed to be eigenstates of some local Hamiltonian $H$.  For 1D CAs, this Hamiltonian can be obtained, e.g., using the parent~\cite{Fannes92,Nachtergaele96,PerezGarcia07} or uncle~\cite{Fernandez15} Hamiltonian construction for matrix product states with finite bond dimension.
The existence of a local parent Hamiltonian $H$ implies that short-orbit eigenstates are \emph{exact} low-entanglement eigenstates of the deformed Floquet unitary $U \equiv \exp(-i\lambda H)\, U_{\rm{aut}}$ for arbitrary $\lambda$, even though this unitary operator generically generates nonintegrable quantum dynamics that do not admit an automaton description~\footnote{Deformations of integrable CA dynamics into fully quantum dynamics were also studied in Ref.~\cite{Friedman19}, where the deformations were chosen to maintain integrability.}. As a corollary, the quantum dynamics under $U$ of a CB state belonging to the short orbit is restricted to a dimension-$\ell$ subspace of the full Hilbert space, precluding the possibility of reaching a true infinite-temperature state. The mechanism discussed here is generic and applies to any CA with short orbits, for which these states are locally identifiable.

An illuminating example of this mechanism is provided by the PXP automaton, which is defined on a spin-1/2 chain with the Floquet unitary
\begin{align}
\label{eq:Uaut}
U_{\rm aut} =\!\prod_{j\text{ odd}}e^{-i\, \frac{\pi}{2}\, (PXP)_j}\!\!\prod_{j\text{ even}}e^{-i\, \frac{\pi}{2}\, (PXP)_j},
\end{align}
where $(PXP)_j=P_{j-1}X_jP_{j+1}$, $P_{j}=(1-Z_j)/2$ is a local projector onto spin-down, and $X_j,Z_j$ are Pauli operators on site $j$. The classical update rule of this CA follows from the identity $e^{\pm i\, \frac{\pi}{2}\, (PXP)_j}={\mathsf{Toffoli}}_j(\pm\pi/2)$ where $\mathsf{Toffoli}_{j}(\phi) \equiv 1-P_{j-1}P_{j+1}+e^{i\phi}\, (PXP)_j$ flips spin $j$ only if its neighbors are pointing down.  This CA has a number of notable features.  First, it can be realized approximately using periodically driven arrays of Rydberg atoms, as we discuss below.  Second, $U_{\rm aut}$ conserves the number of nearest-neighbor pairs of up spins; we may thus restrict our analysis to the sector of the Hilbert space with no such pairs, known as the ``Fibonacci subspace," whose dimension grows like $\varphi^L$ where $\varphi\equiv(1+\sqrt{5})/2$ is the Golden Ratio.  Third, as discussed in the Supplementary Material (SM)~\cite{SM}, the PXP automaton is integrable,
and its dynamics can be described in terms of interacting chiral quasiparticles; a thorough analysis of these classical dynamics has recently been presented in Ref.~\cite{Wilkinson20}. Finally, $U_{\rm aut}$ cannot be recast in the form $e^{-iH_{\rm eff}T}$ for a local Hermitian $H_{\rm eff}$, as can be seen by applying the Baker-Campbell-Hausdorff formula to Eq.~\eqref{eq:Uaut}. The PXP automaton thus provides an example of a Floquet system without a static analog.

The PXP automaton~\eqref{eq:Uaut} supports several short orbits with constant length. Of these, we will focus primarily on the $\ell=3$ ``vacuum orbit," which is defined for even $L$ as $\ket{\Omega}\rightarrow\ket{\mathbb{Z}_{2}^{2}}\rightarrow\ket{\mathbb{Z}_{2}^{1}}\rightarrow\ket{\Omega}$ (up to global phase factors), where the vacuum state $\ket{\Omega}=\ket{0\dots0}$ in the occupation number basis $n_j=(1+Z_j)/2$, while the two N\'eel states $\ket{\mathbb{Z}^1_2}=\ket{10\dots}$ and $\ket{\mathbb{Z}^2_2}=\ket{01\dots}$. Chiral quasiparticles in the PXP automaton are realized as domain walls between configurations in this orbit, hence its designation as a ``vacuum."
We now deform the PXP automaton to a new Floquet unitary $U=U^\prime\, U_{\rm{aut}}$, where $U^\prime$ is any locality-preserving unitary operator that does not mix the vacuum subspace $\{\ket{\Omega},\ket{\mathbb{Z}^1_2},\ket{\mathbb{Z}^2_2}\}$ with other states in the Hilbert space. An example is given by the Hamiltonian evolution operator
\begin{align}
\label{eq:Uprime}
U^\prime\!=\!\exp[-i \lambda\sum_j (-1)^jP_{j-1}\!(X_{j}X_{j+1}\!+\!Y_{j}Y_{j+1})\!P_{j+2}].    
\end{align}
$U^\prime$ acts trivially on any state in the vacuum sector and hence preserves the area-law entangled automaton eigenstates $\ket{\varepsilon_{3,p};\Omega}$ [see Eq.~\eqref{eq:eigenstate}] with quasienergy $0,\pm 2\pi/3$ and the associated threefold-periodic dynamics of CB states in the vacuum sector. 

\begin{figure}[t!]
\begin{center}
\includegraphics[width=\columnwidth]{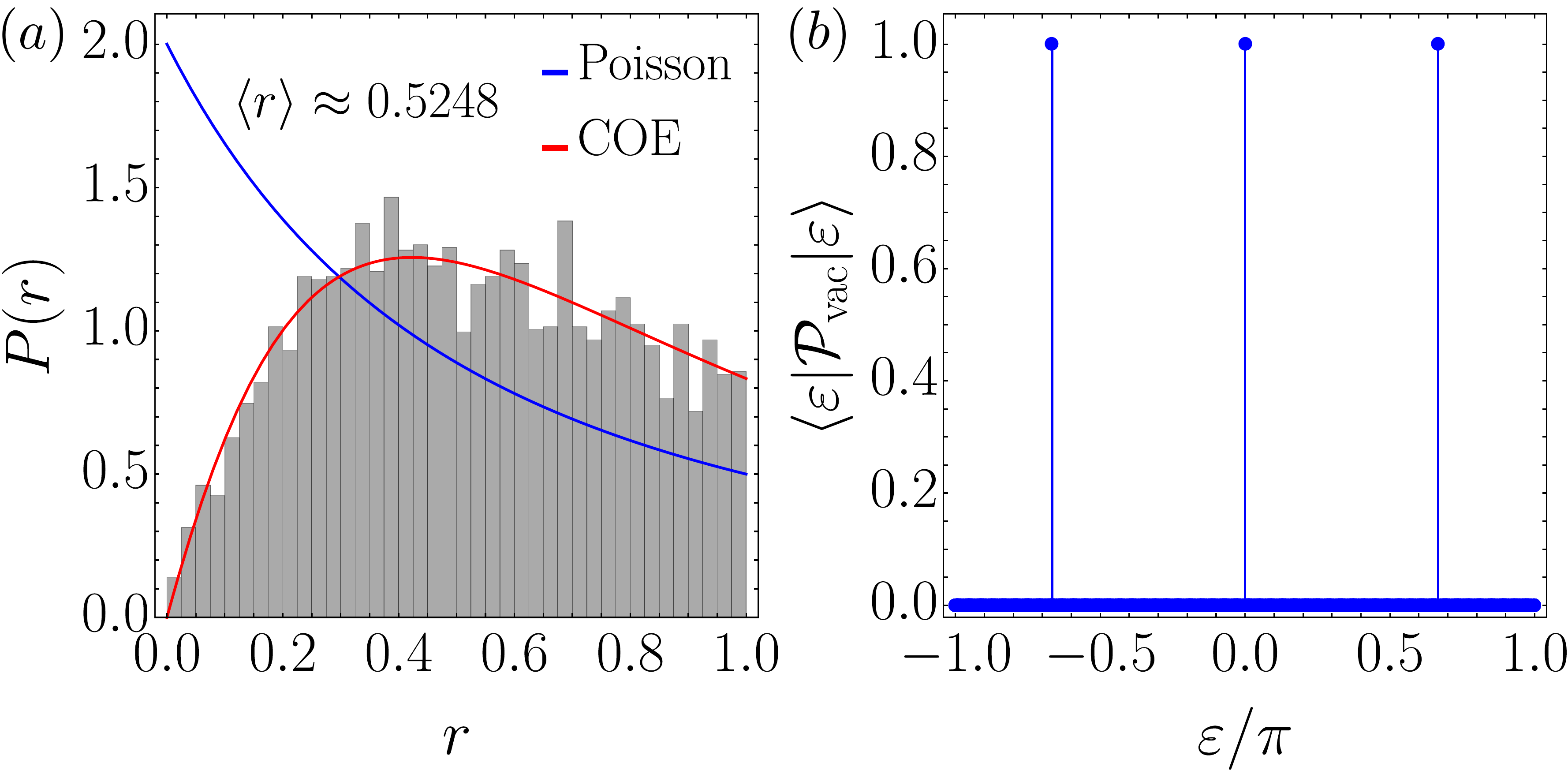}\\
\caption{
Weak ergodicity breaking in a nonintegrable deformation \eqref{eq:Uprime} of the PXP automaton~\eqref{eq:Uaut}.  Data are from ED at $L=24$ in the zero momentum sector of the Fibonacci subspace (8680 states), with $\lambda=0.645\pi/4$.  (a) Distribution of the ratio $r$ of consecutive quasienergy spacings.  The distribution fits the COE of random matrix theory, for which $\langle r\rangle=0.5269$~\cite{D'Alessio14}. (b) Vacuum automaton orbit weight as a function of quasienergy.  Three eigenstates confined to this automaton orbit appear at quasienergies $0,\pm2\pi/3$.
}
\label{fig:deformed-pxp}
\end{center}
\end{figure}

\begin{figure*}[t]
\begin{center}
\includegraphics[width=.85\textwidth]{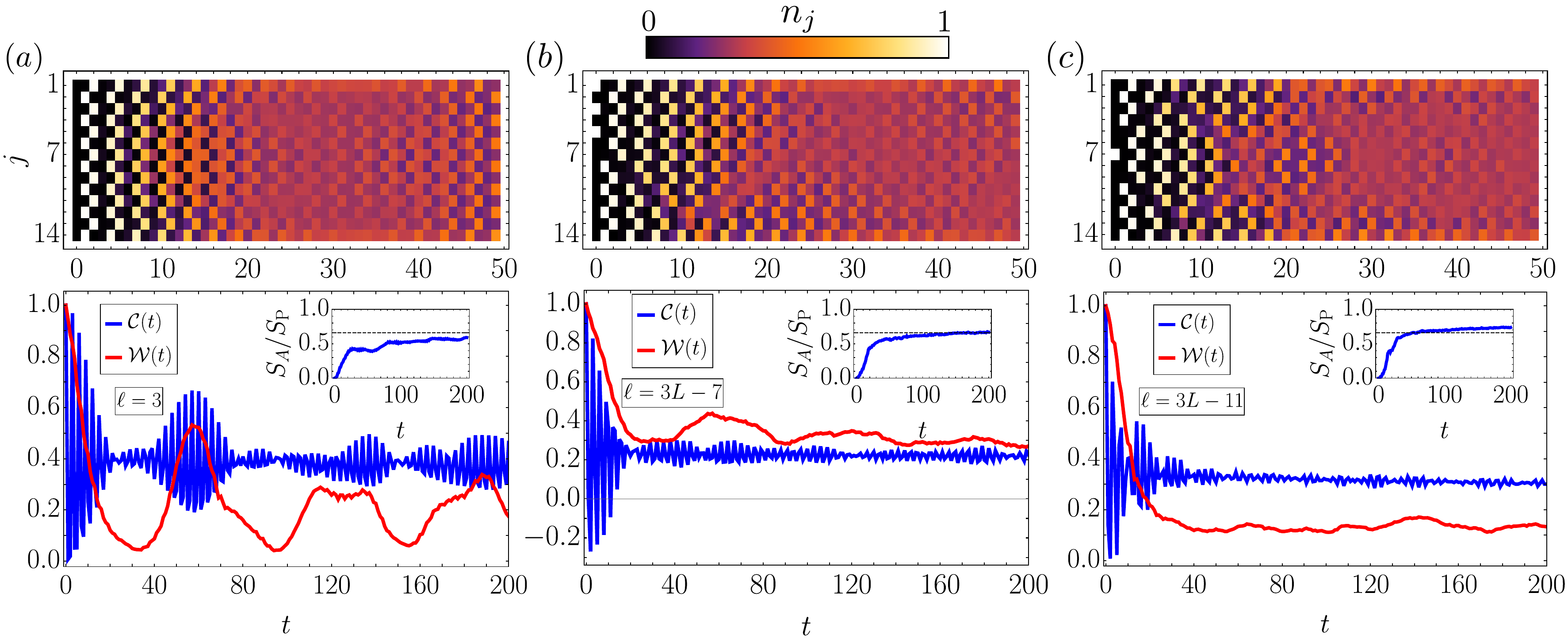}\\
\caption{
Quantum dynamics of three different initial states (a)--(c) under the Ising Hamiltonian \eqref{eq:H} at $L=14$ (OBC) with $h_0T=\pi$ and $V_1=12h_0$, keeping interactions $V_r = V_1/(r)^6$ for $r=1,2,3$.  Top panels show dynamics of the local occupation number $n_j$; bottom panels show dynamics of the autocorrelator $\mathcal C(t)$ and the orbit weight $\mathcal W(t)$ (see text for definitions), and insets show dynamics of the entanglement entropy $S_A$ for half of the chain in units of $S_{\rm P}=(L/2)\ln 2-1/2$.
}
\label{fig:dynamics}
\end{center}
\end{figure*}

Despite acting simply within the vacuum subspace, the deformed unitary $U^\prime\, U_{\rm aut}$ has a complicated action on other CB states and generates nonintegrable dynamics for generic $\lambda$.  This is consistent with numerical data from exact diagonalization (ED) in Fig.~\ref{fig:deformed-pxp}(a), which plots the distribution of the ratio of consecutive quasienergy level spacings~\cite{Oganesyan07,Pal10,D'Alessio14}. The data fit well to random-matrix-theory predictions based on the circular orthogonal ensemble (COE)~\footnote{Note that the COE is the correct random matrix ensemble as $U = U'\, U_{\rm aut}$ has a particle-hole symmetry: $U'(\mathcal C\mathcal M)U(\mathcal M\mathcal C) (U')^\dagger = U^\dagger$, where $\mathcal M$ is a bond-centered mirror reflection and $\mathcal C = \prod_i Z_i$.}, a strong indicator of nonintegrability. Nevertheless, as shown in Fig.~\ref{fig:deformed-pxp}(b), there are three eigenstates in the spectrum with unit weight on the vacuum subspace, as measured by the projector $\mathcal P_{\rm vac}=\sum^2_{k=0}U^k_{\rm aut}\ket{\Omega}\bra{\Omega}(U^\dagger_{\rm aut})^k$, confirming the presence of area-law eigenstates in a nonintegrable Floquet system. This finding strongly contradicts the ETH prediction that all eigenstates are volume-law entangled owing to the nonconservation of energy and consequent lack of a well-defined ground state.

We note in passing that the above construction does not rely on the integrability of the PXP automaton, only on the fact that it has short orbits.  Similar constructions can thus be applied to nonintegrable CAs as well, yielding atypical low-entanglement eigenstates for deformations away from the automaton limit. We do this for a particular nonintegrable CA in the SM~\cite{SM}.

The above discussion demonstrates that quantum deformations of CAs serve as a paradigmatic source of atypical low-entanglement eigenstates and weakly nonergodic quantum dynamics. Although there can be many such deformations for a given CA, they must be chosen to preserve a short-orbit subspace, making them somewhat fine-tuned (albeit less so than the underlying CA). In the SM~\cite{SM}, we argue by generalizing the results of Ref.~\cite{Lin19} for static Hamiltonians that the thermalization time for a \emph{generic} local deformation $e^{-i\lambda H}U_{\rm aut}$ of a 1D CA with $\lambda\ll 1$ can be bounded from below by $t_*\sim \lambda^{-1/2}$. Thus, the thermalization time can be made parametrically large for sufficiently small perturbations.

\emph{\textbf{Emergent Automaton Dynamics with Rydberg Atoms:}}  
 We now provide a specific example of how to realize quantum dynamics near a CA and see evidence of slow thermalization and atypical eigenstates in a laboratory setting. The PXP automaton~\eqref{eq:Uaut} naturally arises in a limit of driven dynamics using Rydberg states (see also Ref.~\cite{Wintermantel20}). We consider a Rydberg-atom quantum simulator like that of Ref.~\cite{Bernien17}, which realizes a long-range quantum Ising model,
\begin{align}
\label{eq:H}
H(t)=\sum_{i<j}V_{i,j}\, n_i\, n_j + \sum_{i} 
h_{i}(t)\, X_i. 
\end{align}
The long-ranged repulsive coupling $V_{i,j}\sim |i-j|^{-6}$ arises due to Van der Waals interactions between excited atoms; we label interaction terms by their range as $V_{r}=V_{i,i+r}$. The transverse field $h_i$ is proportional to the Rabi frequency of a pump pulse that drives atoms between their ground and excited Rydberg states.
We now consider the spatially modulated square-wave drive 
\begin{align}
\label{eq:SquareWave}
h_{j}(t) = h_{0} \frac{1 + (-1)^{j}\,\text{sgn}(\sin\omega t)}{2},
\end{align}
with $\omega=2\pi/T$.  In the SM~\cite{SM}, we show that in the presence of only nearest-neighbor interactions ($V_{j} = 0$ for $j \ge 2$), and when $h_{0}/V_{1} \ll 1$, the stroboscopic dynamics generated by this time-dependent Hamiltonian is given by a Floquet operator (the ``PXP circuit"):
\begin{align}
\label{eq:UF}
U_{\rm F}(\theta) =\!\prod_{j\text{ odd}}e^{-i\, \theta\, (PXP)_j}\!\!\prod_{j\text{ even}}e^{-i\, \theta\, (PXP)_j},
\end{align}
with $\theta\equiv h_{0}T/2$, so that $\mathcal{T}\left\{\exp[{-i\int_{0}^{nT}dt\,H(t)}]\right\} = (U_{F}(\theta))^{n}$.  
We 
distinguish two limits of interest in Eq.~\eqref{eq:UF}.
First, in the limit of a high-frequency drive ($\theta\to0$), $U_{\rm F}$ can be viewed as a ``Trotterization" of the 
dynamics generated by $H_{\rm PXP}=\sum_j(PXP)_j$, which feature persistent revivals characteristic of QMBS~\cite{Turner18b}. 
Second, when $\theta=\pm \pi/2$ mod $2\pi$, $U_{\rm F}$ is equivalent to the PXP automaton~\eqref{eq:Uaut}.
In fact, we show in the SM~\cite{SM} that, in the presence of arbitrarily long-ranged $V_r\sim V_1/r^6$, this CA regime supports a \emph{family} of automaton dynamics whose classical update rule conditions a spin flip at site $j$ on the state of its further neighbors, depending on the interaction strength $V_1$.

\emph{\textbf{Numerical Results:}} 
We now numerically study the dynamics of the time-dependent Hamiltonian (\ref{eq:H}) near the automaton point.
We use open boundary conditions (OBC) and the realistic interaction strength $V_1=12h_0$~\cite{Bernien17}, keeping interactions $V_r=V_1/r^6$ up to $r=3$, and further set $\theta=\pi/2$.  Note that, even for $\theta=\pi/2$, the model \eqref{eq:H} realizes \emph{perturbed} automaton dynamics due to both the presence of $V_2=V_1/2^6\approx 0.2h_0$ and subleading corrections to Eq.~\eqref{eq:UF}~\cite{SM}.
Nevertheless, we will find signatures of both slow dynamics for initial states in the vacuum orbit and quasiparticles from the underlying CA.

Our numerical results for $L=14$ are shown in Fig.~\ref{fig:dynamics}.  The top panels in Fig.~\ref{fig:dynamics} show the dynamics of the local occupation number $n_j$ at relatively early times for three initial CB states: the vacuum state $\ket{\Omega}$ (a), a one-quasiparticle state (b), and a two-quasiparticle state (c), which belong to automaton orbits with $\ell=3,3L-7,$ and $3L-11$, respectively.  The proximate automaton dynamics is plainly visible in these plots, including the existence of chiral quasiparticles.  Most signatures of quasiparticles wash out by $t\sim50\, T$; however, the dynamics of the short-orbit initial state $\ket{\Omega}$ remains coherent at this timescale.  

We study longer-time quantum dynamics in the bottom panels of Fig.~\ref{fig:dynamics}, which show the local spin autocorrelator $\mathcal C(t)=\sum^L_{i=1}\langle Z_i(t)Z_i(0)\rangle/L$, the automaton orbit weight $\mathcal W(t)$, and the bipartite entanglement entropy $S_A(t)$ for one half of the chain.  The autocorrelator $\mathcal C(t)$ measures the average likelihood of a spin to return to its initial CB state at time $t$, and shows a clear dependence on the initial state.  The coherent dynamics of the vacuum initial state $\ket{\Omega}$ manifests itself as period-3 oscillations in $\mathcal C(t)$, which remain sharply visible even at $t=200\, T$.  Period-3 oscillations are also visible at early times for the one- and two-quasiparticle initial states, as much of the system locally resembles the vacuum, but these decohere by $t\sim 40\, T$.
We define the automaton orbit weight $\mathcal W(t)=\langle \mathcal P_{\rm orbit}(t)\rangle$, where $\mathcal P_{\rm orbit}=\sum^{\ell-1}_{k=0}U^k_{\rm aut}\ket{\psi_0}\bra{\psi_0}(U^\dagger_{\rm aut})^k$ is the projector onto the length-$\ell$ orbit of the state $\ket{\psi_0}$. $\mathcal W(t)$ tracks the fraction of the time-evolved state that remains in its original automaton orbit; we see in Fig.~\ref{fig:dynamics} that the vacuum initial state experiences pronounced revivals in $\mathcal W(t)$ that coincide with periods of enhanced coherence in $\mathcal C(t)$, while the other initial states exhibit a fast initial drop in $\mathcal W(t)$ followed by gradual decay that persists beyond the times shown.  The insets of these plots show the dynamics of $S_A$ in units of $S_{\rm P}=(L/2)\ln 2-1/2$, Page's result for a random state of $L$ qubits~\cite{Page93}; the dashed line indicates the Page value for the Fibonacci subspace, which is approximately closed under the dynamics.  We see that real-space entanglement develops rapidly as expected for a quantum-chaotic system, although it develops comparatively slowly for the vacuum initial state.

\begin{figure}[t!]
\begin{center}
\includegraphics[width=\columnwidth]{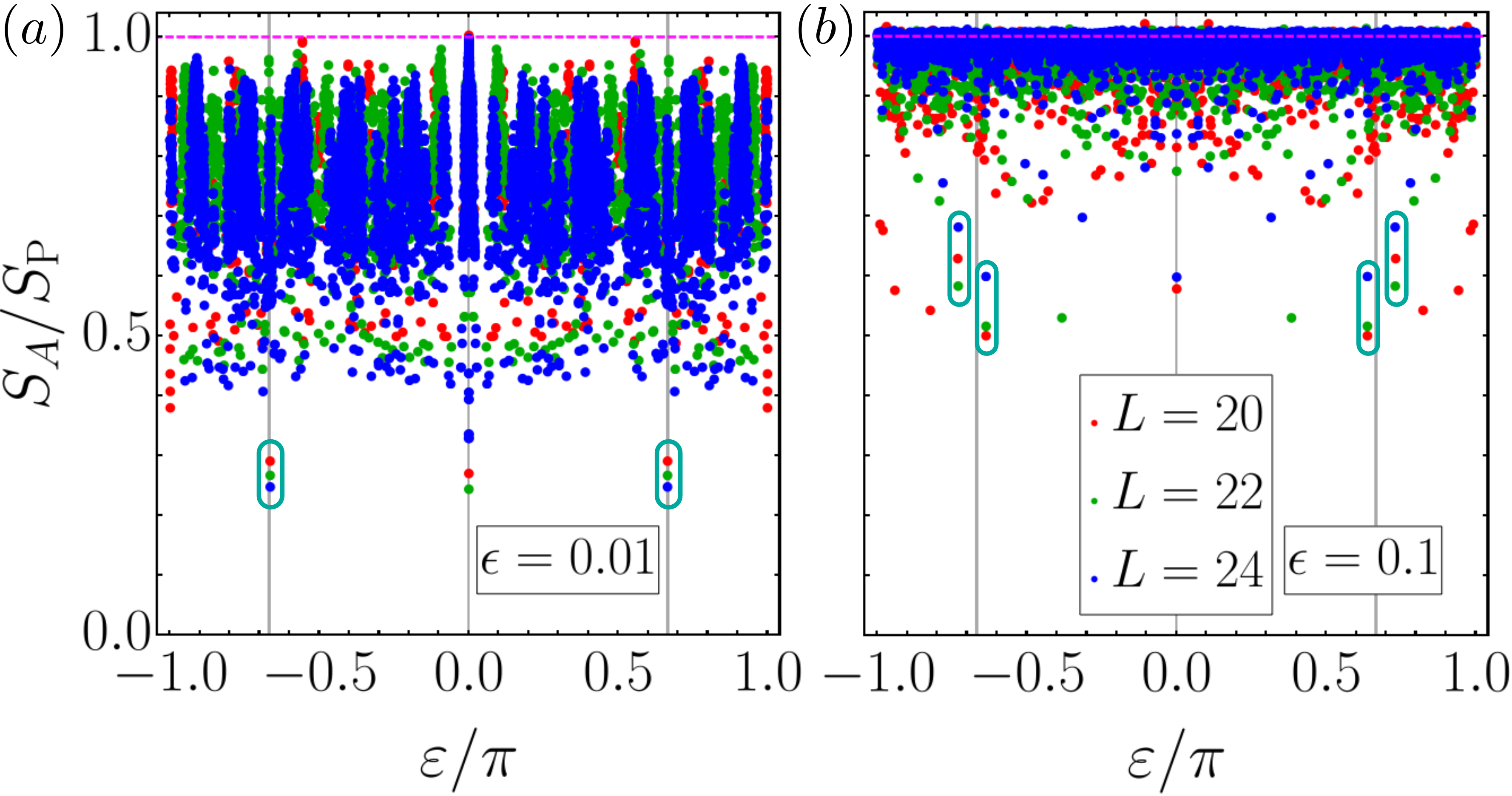}\\
\caption{
Half-chain entanglement entropy as a function of quasienergy for the PXP circuit~\eqref{eq:UF} with $\theta=\pi/2(1+\epsilon)$ and $\epsilon=0.01$ (a) and $0.1$ (b). Data for $L=20,22,24$ in the zero-momentum sector of the Fibonacci subspace are shown. Entanglement outliers associated with the vacuum automaton orbit are circled in teal.
}
\label{fig:pxp-entanglement}
\end{center}
\end{figure}


The enhanced coherence of the vacuum state's dynamics suggests that the perturbed PXP automaton has low-entanglement eigenstates descending from the CA eigenstates $\ket{\varepsilon_{3,p};\Omega}$ that retain substantial weight on the $\ell=3$ vacuum orbit, analogous to QMBS states in the static PXP Hamiltonian $H_{\rm PXP}$~\cite{Bernien17,Turner18a}. We now numerically study these eigenstates of the PXP circuit \eqref{eq:UF} with $\theta=(\pi/2)(1+\epsilon)$ and $0<\epsilon\ll1$.

Descendants of the CA eigenstate $\ket{\varepsilon_{3,p};\Omega}$ are plainly visible in Fig.~\ref{fig:pxp-entanglement}, which plots the half-chain entanglement entropy $S_A$ [now in units of $S_{\rm P}=(L/2)\ln\varphi-1/2$ since we work in the Fibonacci Hilbert space] as a function of the quasienergy $\varepsilon$ for $L=20,22,24$ and $\epsilon=0.01$ (a) and $0.1$ (b).  For $\epsilon=0.01$, there is a broad distribution of entanglement entropy inherited from that of the proximate CA. Clear low-entanglement outliers (circled in teal) are visible near $\varepsilon=\pm 2\pi/3$ (vertical grey lines); we focus on these eigenstates as the model \eqref{eq:UF} has a large quasienergy degeneracy near $\varepsilon=0$~\cite{Schecter18} that complicates the identification of the eigenstates of interest.  The entanglement outliers near $\varepsilon=\pm 2\pi/3$ persist for much larger values of $\epsilon$, as shown in Fig.~\ref{fig:pxp-entanglement}(b).  At $\epsilon=0.1$, entanglement outliers near $\varepsilon=\pm2\pi/3$ remain prominent [note that $\epsilon=0.5$ yields the maximum distance from any nongeneric point in parameter space, see Eq.~\eqref{eq:UF}].  Moreover, an additional set of slightly higher-entanglement outliers appears near $\varepsilon=\pm 2\pi/3$; we have verified by calculating $\langle\mathcal P_{\rm vac}\rangle$ that these outliers also descend from $\ket{\varepsilon_{3,p};\Omega}$.

In the SM~\cite{SM}, we analyze the scaling with $\epsilon$ and $L$ of the vacuum orbit weight $\langle\mathcal P_{\rm vac}\rangle$ and the bipartite entanglement entropy $S_A$ for the entanglement outlier closest to $\varepsilon=-2\pi/3$. We find that, for small $\epsilon$, $\langle\mathcal P_{\rm vac}\rangle\sim(1+\epsilon)^{-L}$ while $S_A\sim \epsilon L$. Both approximate scaling forms are consistent with these eigenstates residing in an effective Hilbert space of dimension $\sim(1+\epsilon)^L$. This effective Hilbert space is exponentially smaller than the full Hilbert space dimension $\sim\varphi^L$, confirming the atypical nature of the entanglement outliers at small $\epsilon$. We expect the dynamics shown in Fig.~\ref{fig:dynamics} to reflect a similar finite-size dependence of the revival amplitude at late times. The small system sizes accessible to ED lead to uncertainty about the persistence of these properties as $L\to\infty$, similar to the case of QMBS in the PXP model~\cite{Turner18b,Iadecola19,Lin19}. Nevertheless, these numerical results show that atypical eigenstates and weakly nonergodic dynamics can be obtained in finite systems by perturbing the PXP automaton without fine tuning.

We have shown that nonthermal eigenstates of classical CAs can remain as exact eigenstates of Floquet dynamics that exhibit otherwise quantum-chaotic behavior, and that certain signatures of this ergodicity-breaking can emerge in experiments in Rydberg atom quantum simulators.  A number of directions related to these ideas and their realizations in experiments remain to be explored.  Certain 1D CAs appear to have logarithmic entanglement entropy~\cite{Sarang_Automaton1}, reminiscent of that of a critical one-dimensional quantum system, though some of these states cannot be realized as eigenstates of static Hamiltonians (e.g. due to the presence of chiral quasiparticles in the classical CA dynamics).  Signatures of these nonthermal states in Floquet dynamics that are quantifiably close to the CA point are not understood.  Furthermore, nonthermal eigenstates in higher-dimensional CAs can potentially be used to understand dynamics in more exotic constrained quantum systems, some of which can exhibit fractionalization  \cite{celi2019emerging}.   Finally, connections between continuous-time quantum dynamics and rare orbits in continuous, classically chaotic systems deserve greater study. 

\emph{Acknowledgments:} We thank the organizers of the 2019 Les Houches Summer School, ``New Developments in Topological Condensed Matter," where this work was initiated.  This work is supported by Iowa State University startup funds (TI) and the Harvard Society of Fellows (SV).

\bibliography{refs}

\pagebreak

\appendix

\renewcommand{\thefigure}{S\arabic{figure}}
\setcounter{figure}{0}

\begin{center}
    \textbf{Supplemental Material for ``Nonergodic Quantum Dynamics from Deformations of Classical Cellular Automata"}
\end{center}

\section{Short Orbits in Translationally-Invariant Classical Cellular Automata}
\label{sec:CA_short_orbits}
In this section, we show by construction that translationally-invariant classical CAs  always admit at least one ``short" orbit. Consider a 1D cellular automaton defined by the unitary operator $U_{\mathrm{aut}}$, and suppose that these dynamics are invariant under translation $T_{n}$ by $n$ lattice sites. Since $[T_n,U_{\rm aut}]=0$, a translationally-invariant CB state $\ket{\psi}$ will always satisfy $T_{n}\ket{\psi(t)} = \ket{\psi(t)}$ under time evolution by the CA.  However, there are only $2^{n}$ such CB states, so that $\ket{\psi}$ can only belong to an orbit of length $\ell \le 2^{n}$, which is constant in the thermodynamic limit.  This argument generalizes straightforwardly to translationally-invariant CAs in higher dimensions by considering computational basis states that are invariant under all elementary translations. We conclude that, for any translationally-invariant CA, there are always orbits whose length does not grow with the system size.

\section{Quasiparticles in the PXP Automaton}
\label{sec:Quasiparticles}

\begin{figure}[b!]
    \includegraphics[width=.5\columnwidth]{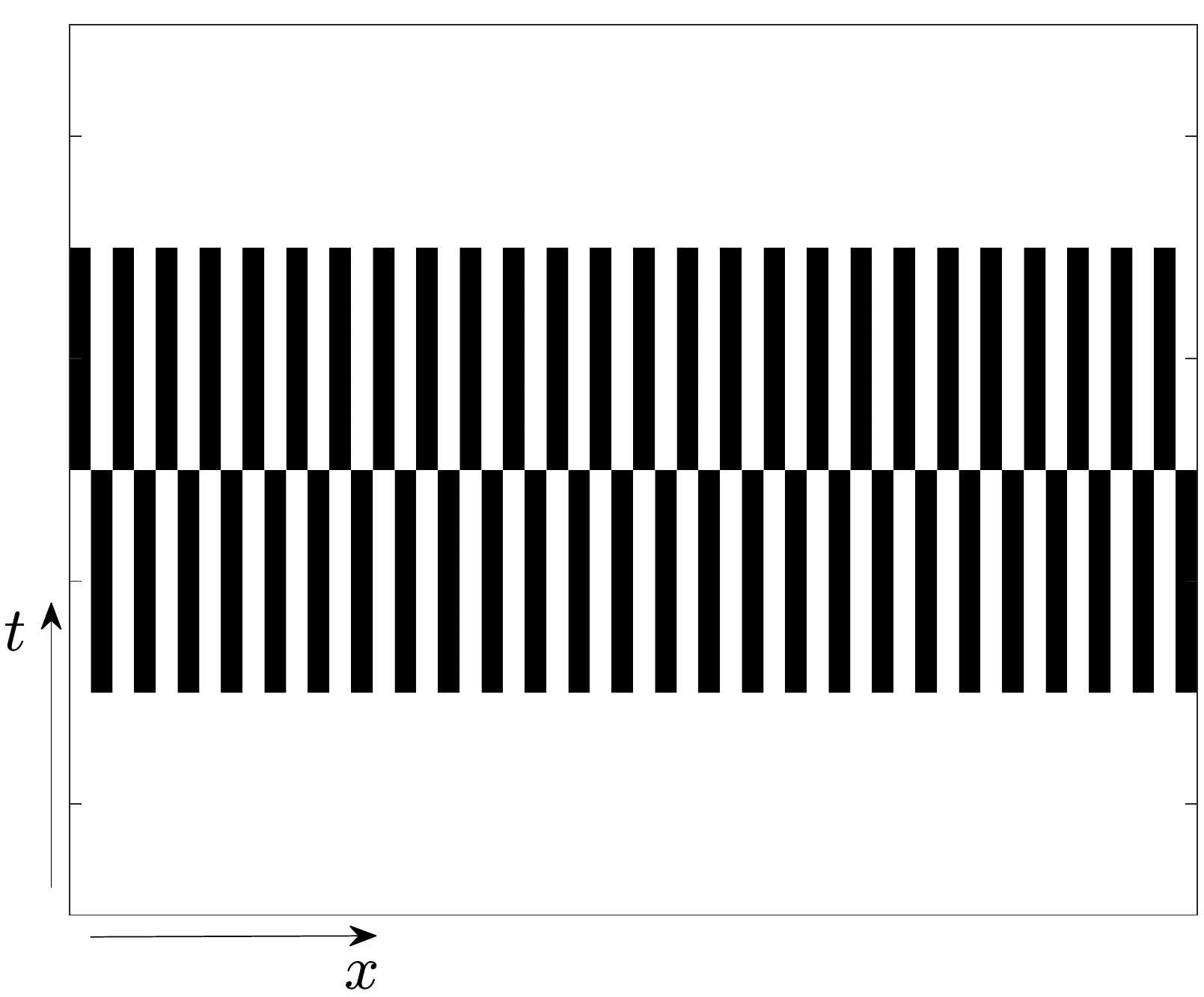} 
    \caption{The dynamics of the computational basis states in the length-3 orbit of the PXP automaton.  Here black and white regions correspond to occupied and unoccupied sites.}
    \label{fig:l3_orbit}
\end{figure}

We now study the quasiparticles in the classical PXP automaton, as defined in the main text.  Given the constant-length orbits formed by the computational basis (CB) states under the PXP automaton, it is natural to consider the behavior of quasiparticle states, which are formed of ``domain walls" between different kinds of states which belong to short orbits.

\begin{figure*}
$\begin{array}{cccc}
\includegraphics[width=.5\columnwidth]{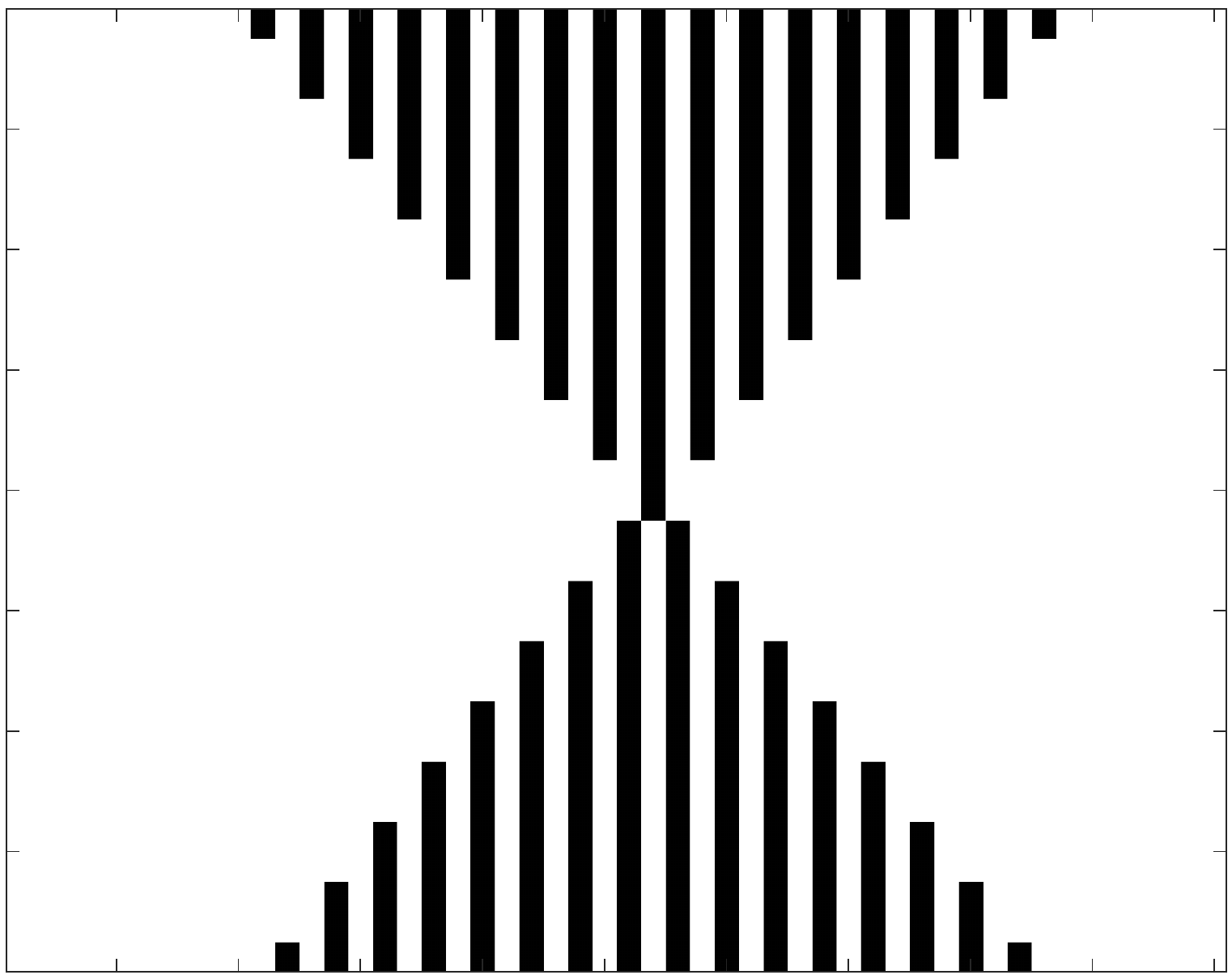} & 
\includegraphics[width=.5\columnwidth]{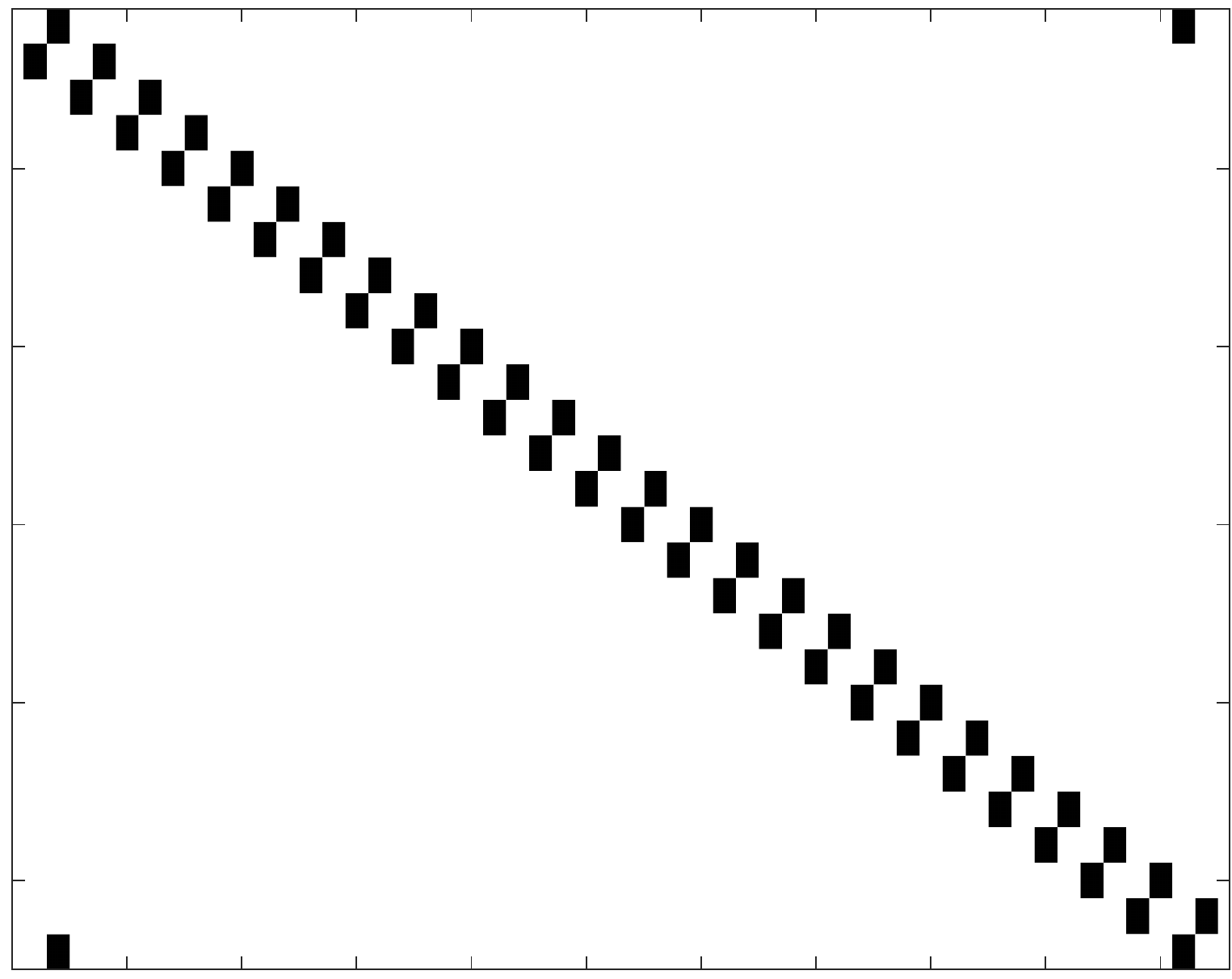} & 
\includegraphics[width=.5\columnwidth]{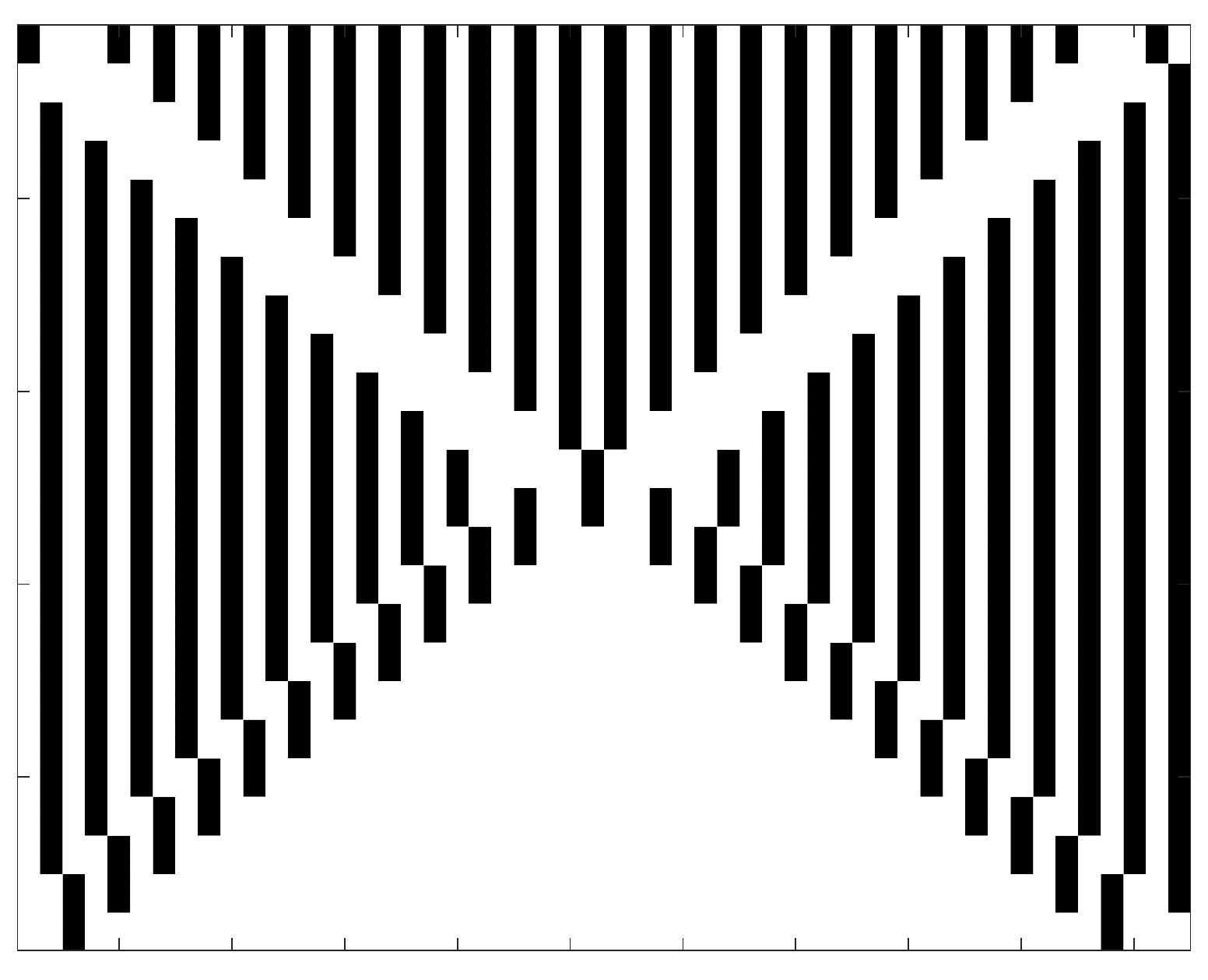}
& \includegraphics[width=.5\columnwidth]{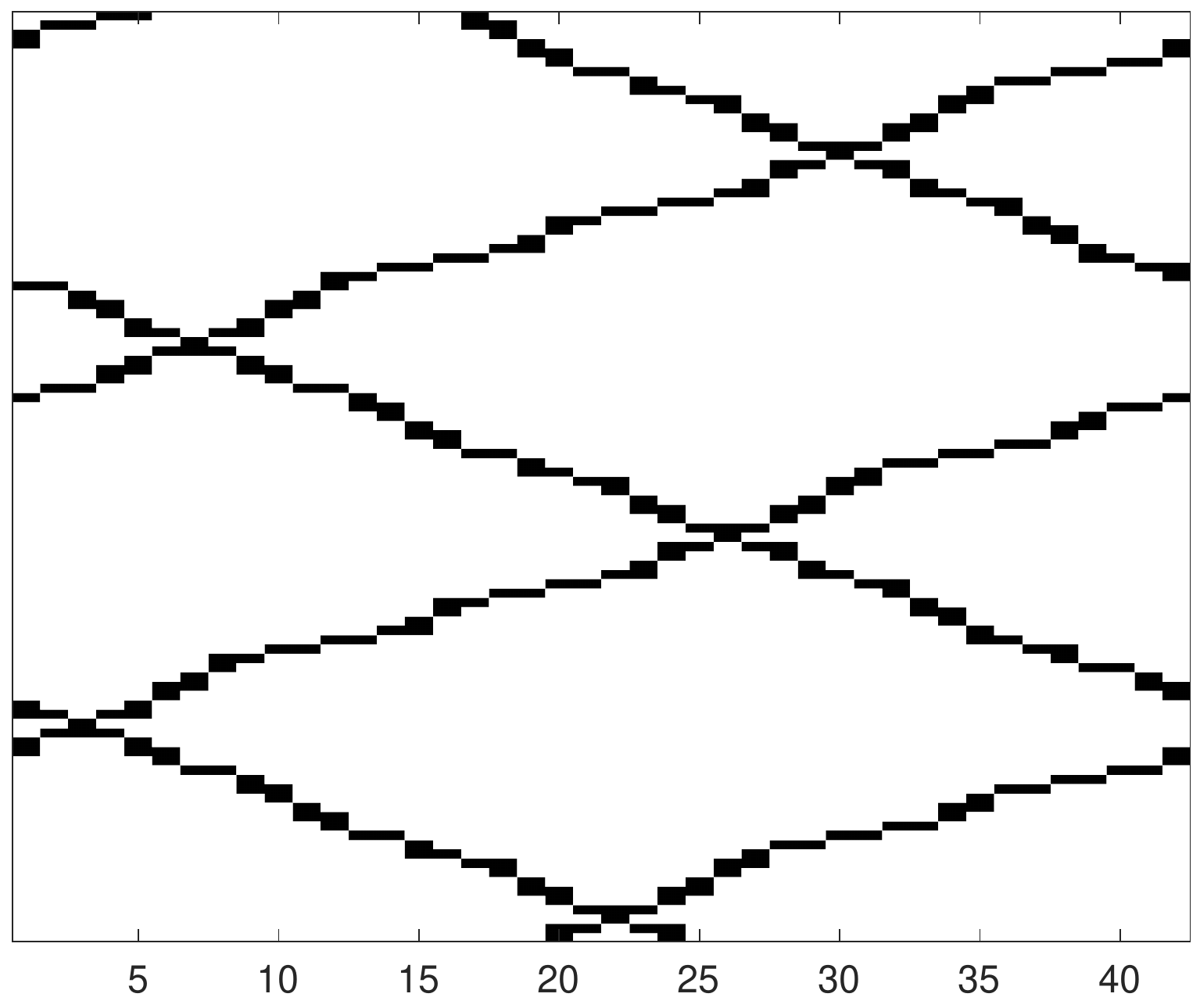}
\\
\text{(a)} & \text{(b)} & \text{(c)} & \text{(d)}
\end{array}$
\caption{In (a)-(c) are shown the dynamics of different states formed by juxtaposing states in the $\ell=3$ vacuum orbit.  These ``domain walls" propagate as local quasiparticles.  In all three figures, and for simplicity of presentation, the dynamics generated by the PXP automaton are shown at three-timestep intervals.  Certain quasiparticles, such as the one in (b) propagate \emph{chirally}.  The classical chaos diagnostic $G(x,t) \equiv ||n_{1}(x,t) -  n_{2}(x,t)||$  is plotted in (d) for an $L = 42$ site system, as described in the text.  This correlator remains spatially-localized at all times, indicating that the dynamics are integrable.}
\label{fig:quasiparticles}
\end{figure*}

A simple example is given by the CB states in the $\ell=3$ orbit, which are $\ket{\Omega}$, $\ket{\mathbb{Z}_{2}^{2}}$, and $\ket{\mathbb{Z}_{2}^{1}}$.  The evolution of these states under application of the PXP automaton is shown in Fig.~\ref{fig:l3_orbit} for $L = 42$ sites, with white and black regions denoting unoccupied and occupied states of the sites, respectively.  Quasiparticle states are then formed by juxtaposing these CB states; a simple example is given in Fig.~\ref{fig:quasiparticles}(a), in which the initial state consists of a region which locally looks like the N\'{e}el state, and one that looks like the fully-polarized $\ket{\Omega}$ state in an $L = 50$ site system.  For added clarity, the  evolution of the initial state is only shown at every third timestep (i.e. the dynamics generated by $U_{\mathrm{aut}}^{3}$).  It is evident that the domain walls between two states within this $\ell=3$ short orbit propagate as left- and right-moving quasiparticles.  

Other quasiparticles can be constructed in a similar fashion.   Fig.~\ref{fig:quasiparticles}(b) shows a  \emph{chirally} propagating quasiparticle, which is formed from an initial state $\ket{0\ldots0 10010\ldots 0}$, which can be thought of as a bound-state of two of the domain walls that are formed within the $\ell=3$ short-orbit CB states.  The nontrivial interactions between these quasiparticles are evident in Fig. \ref{fig:quasiparticles}(c), which shows that certain quasiparticle types can ``change identity" following a scattering event.

Finally, while these examples are very suggestive that the PXP automaton dynamics are integrable, a more formal test of the integrability of these dynamics is given by studying the correlation function $G(x,t)\equiv || n_{1}(x,t) - n_{2}(x,t)||$, where $n_{1,2}(x,0)$ are two CB states in the Fibonacci subspace that initially only differ locally.  Integrability may be diagnosed by the fact that this correlation function remains localized in space as it evolves \cite{Classical_OTOC}, as is evident in the plot of $G(x,t)$ in Fig.  \ref{fig:quasiparticles}(d).  Here, two states are chosen which are identical away from the origin. Near the origin, the two states differ at two sites; the first state is given by $\ket{\cdots 000010000\cdots}$, while the second state is given by $\ket{\cdots001010100\cdots}$.

\section{Example of a Nonintegrable Automaton with Short Orbits}

In this Appendix we provide an example of a nonintegrable CA with short orbits and demonstrate that the deformation procedure outlined in the main text can be used to obtain nonintegrable quantum dynamics that preserves one of these orbits.  The CA we consider is defined by the Floquet unitary
\begin{align}
\label{eq:NonintegrableAut}
    U_{\rm aut} = U_1(\pi)\,U_3(\pi)\,U_2(\pi)\, U^\mathrm{PXP}_{\rm aut},
\end{align}
where $U^\mathrm{PXP}_{\rm aut}$ is the Floquet unitary for the PXP automaton~\eqref{eq:Uaut} and 
\begin{widetext}
\begin{align}
\label{eq:UnDef}
U_n(\theta)= \exp\Big[-i\theta\sum^{L/3-1}_{j=0} P_{3j+n-1}\big(X_{3j+n}X_{3j+n+1}+Y_{3j+n}Y_{3j+n+1}\big)P_{3j+n+2}\Big]
\end{align}
\end{widetext}
for $n=1,2,3$. We take $L=0$ mod 6 and periodic boundary conditions for simplicity.

\begin{figure}[b!]
\begin{center}
\includegraphics[width=\columnwidth]{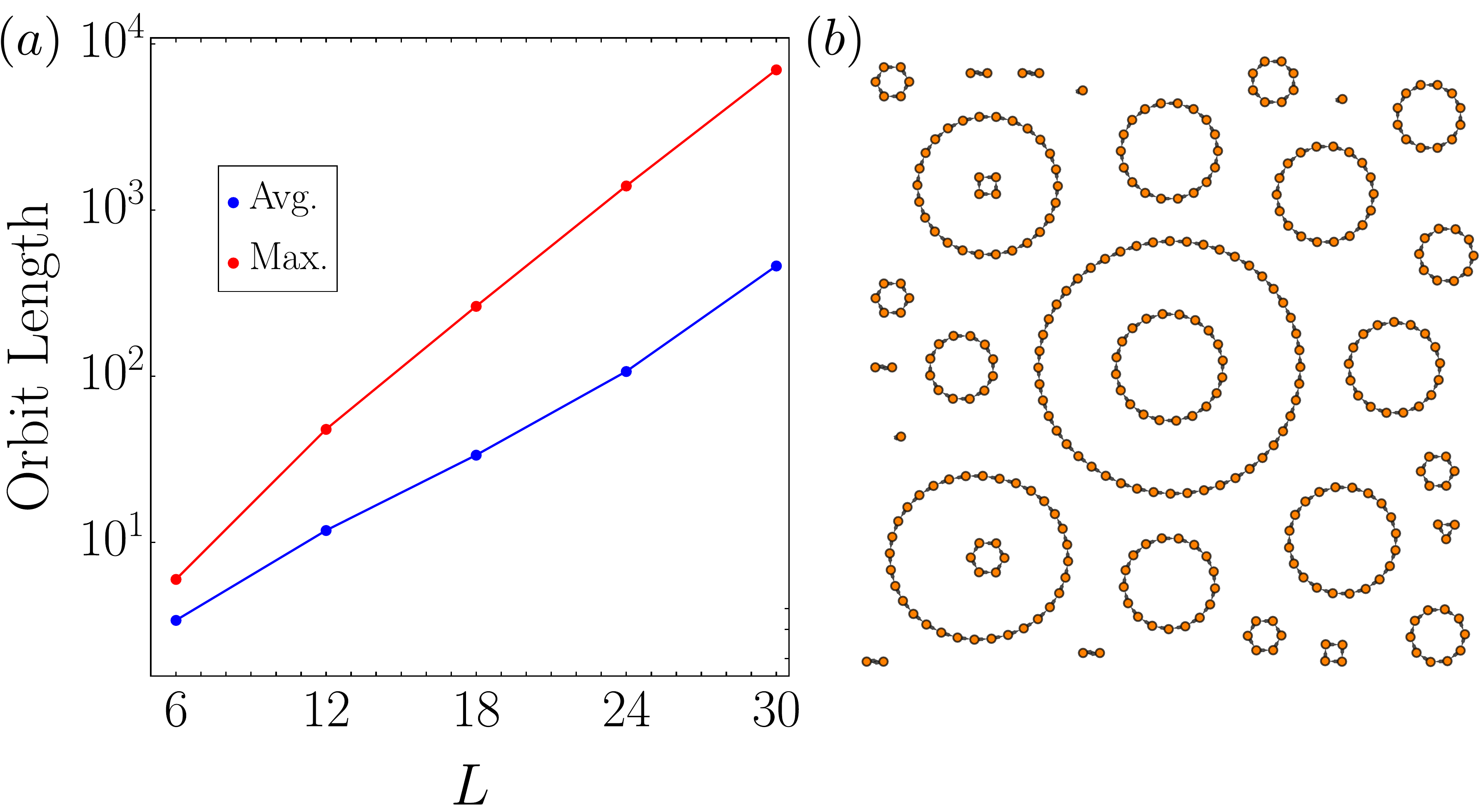}\\
\caption{
Classical orbits in the nonintegrable automaton~\eqref{eq:NonintegrableAut}. (a) Semilog plot of average and maximum orbit lengths as a function of $L$. Clear exponential scaling, a signature of classical chaos, is visible. (b) Graphical visualization of the action of the nonintegrable automaton~\eqref{eq:NonintegrableAut} on CB states.  Orbits of various lengths are visible, including short orbits with $\ell=1$ and $\ell=3$, discussed in this Appendix.
}
\label{fig:nonintegrable}
\end{center}
\end{figure}

One way to verify that Eq.~\eqref{eq:NonintegrableAut} defines a (classically) nonintegrable automaton evolution is by considering the scaling with $L$ of a typical orbit length.  In Fig.~\ref{fig:nonintegrable}(a), we see that both the average and maximum orbit lengths scale exponentially with system size. Moreover, calculating the chaos diagnostic $G(x,t) \equiv ||n_{1}(x,t) - n_{2}(x,t)||$ introduced in Appendix \ref{sec:Quasiparticles}, as shown in Fig.~\ref{fig:chaotic_correlator},  confirms the classically chaotic nature of these CA dynamics.   Despite this, the graphical representation of the action of $U_{\rm aut}$ on the CB in Fig.~\ref{fig:nonintegrable}(b) reveals the presence of short orbits.  Indeed, one of these orbits is the vacuum orbit of $U^\mathrm{PXP}_{\rm aut}$, which is the focus of the main text.  We can deduce the existence of this orbit for Eq.~\eqref{eq:NonintegrableAut} by noting that the unitaries $U_n(\theta)$ defined in Eq.~\eqref{eq:UnDef} act trivially on the vacuum subspace $\{\ket{\Omega},\ket{\mathbb{Z}^1_2},\ket{\mathbb{Z}^2_2}\}$, as their Hermitian generators annihilate the CB states spanning it.  (This reasoning also underlies our choice of the operator $U^\prime$ in the main text.)  Another short orbit under this automaton is the $\ell=1$ orbit consisting of the state $\ket{\mathbb Z^5_6}=\ket{000010\dots}$. 

We can now provide a deformation of the nonintegrable automaton $U_{\rm aut}$ that preserves the vacuum orbit while otherwise destroying the automaton dynamics.  The choice of deformation is clear given the fact that $U_n(\theta)$ preserve the vacuum subspace for any $\theta$:
\begin{align}
    U(\theta)=U_1(\theta)\,U_3(\theta)\,U_2(\theta)\, U^\mathrm{PXP}_{\rm aut}.
\end{align}
Like the deformed automaton considered in the main text, the Floquet unitary $U(\theta)$ exhibits nonintegrable quantum dynamics but retains threefold periodic dynamics for CB initial states in the vacuum sector, as well as the associated low-entanglement automaton eigenstates.  We can also deform $U_{\rm aut}$ in a manner that preserves \emph{both} the vacuum orbit and the CB state $\ket{\mathbb Z^5_6}=\ket{000010\dots}$, which forms its own $\ell=1$ orbit.  This is achieved by the Floquet operator $U=U^\prime\, U_{\rm aut}$ with
\begin{align}
U^\prime=\exp[-i \lambda\sum_j P_{j-1}(X_{j}X_{j+5}\!+\!Y_{j}Y_{j+5})P_{j+6}].    
\end{align}
We thus see that a given automaton can be deformed in various ways to preserve different short orbits and combinations thereof, even when the underlying automaton is nonintegrable.

\begin{figure}[t!]
\includegraphics[width=.75\columnwidth]{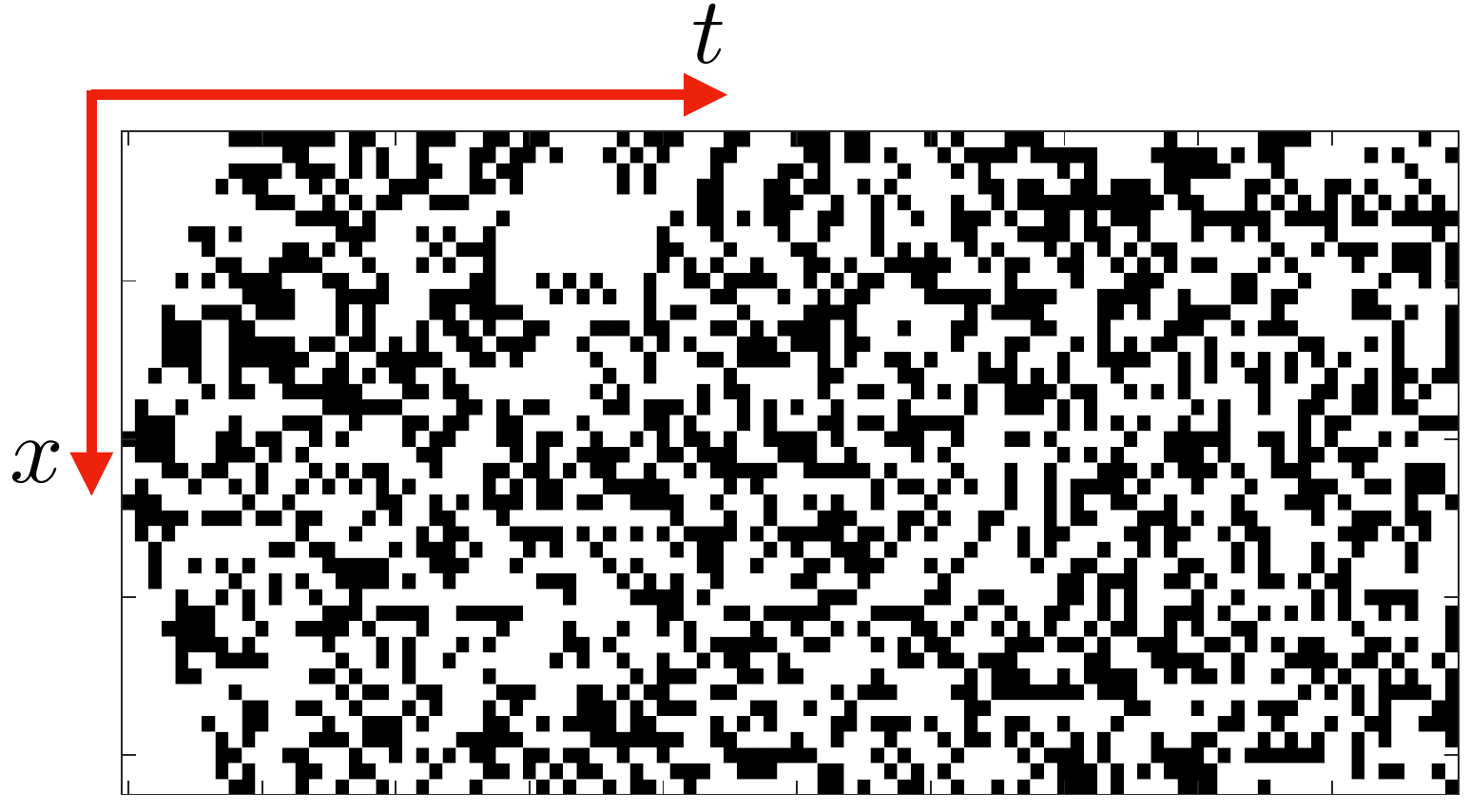}\\
\includegraphics[width=.75\columnwidth]{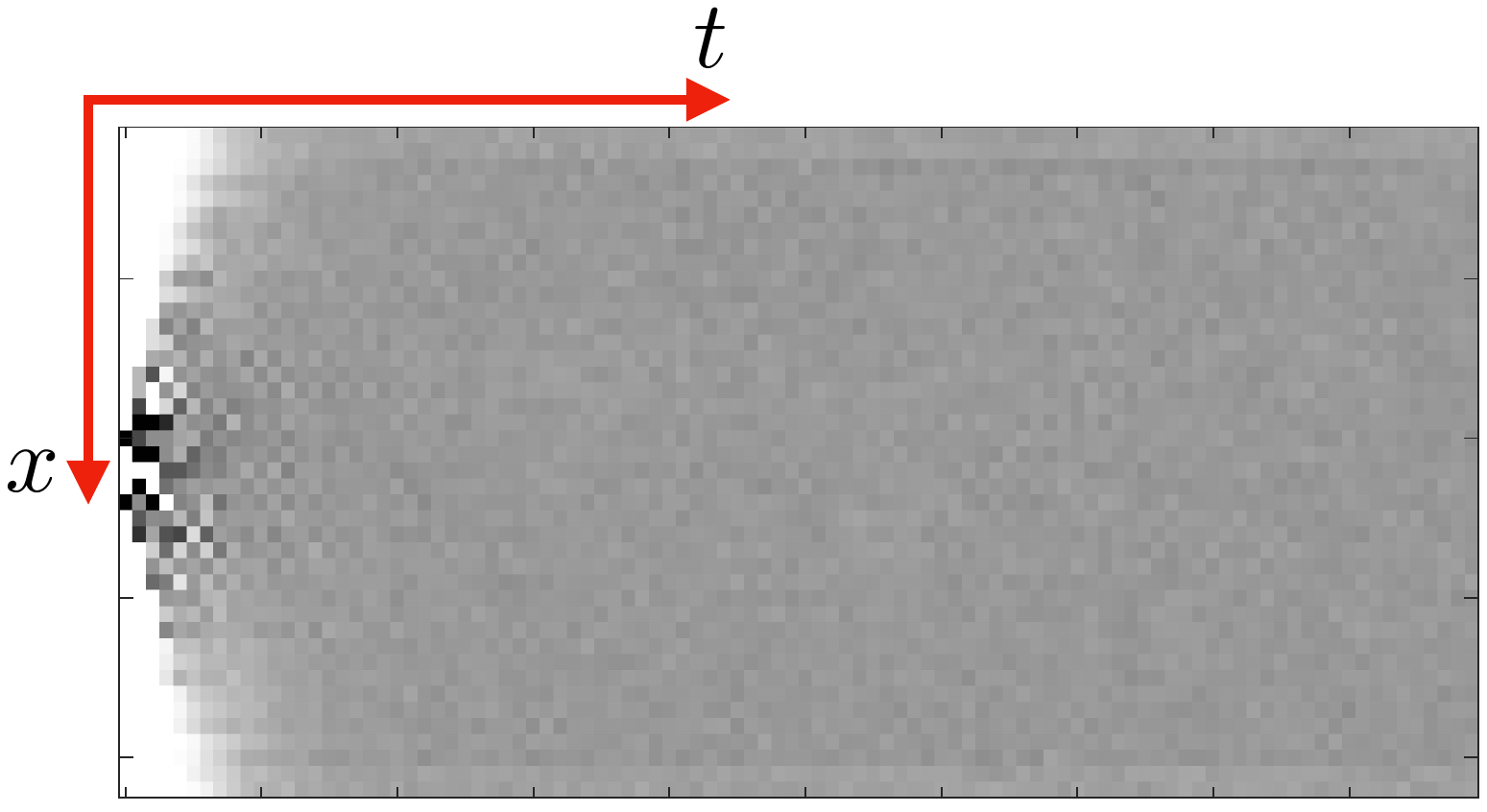}
\caption{A plot of the classical chaos diagnostic $G(x,t)$ for the CA dynamics defined by Eq. (\ref{eq:NonintegrableAut}).  The two initial states $n_{1}(x,0)$ and $n_{2}(x,0)$ are the chosen to be identical away from the origin.  Near the origin, however, the two states differ; the first state is initially given by $\ket{\cdots 000010000\cdots}$ and the second is $\ket{\cdots 001010100\cdots}$.  The growth of  $G(x,t) = ||n_{1}(x,t) - n_{2}(x,t)||$ for a particular choice of these states is shown in the top figure.  This correlation function, averaged over $10^{3}$ choices of the initial states away from the origin, is shown in the bottom figure.  In both cases, the spatial growth of $G(x,t)$ in time provides a signature of the classically chaotic nature of these dynamics.
}
\label{fig:chaotic_correlator}
\end{figure}

\section{Bound on Thermalization Timescale for a Generic Local Perturbation}

Let $U_{\rm aut}$ be a Floquet unitary acting on a 1D chain that generates automaton dynamics, and consider the perturbed unitary $U=e^{-i\, \lambda\, H}\, U_{\rm aut}$, where $0<\lambda\ll 1$ and $H$ is a local Hamiltonian.  We would like to bound the time scale $t_*$ beyond which the atypical features of a short-orbit eigenstate $\ket{\varepsilon}$ of $U_{\rm aut}$ are destroyed by the perturbation.  To do this, we generalize an argument due to Lin et al.~\cite{Lin19}, who derived a bound for the rate of change of a local operator under continuous-time Hamiltonian dynamics. Our generalization applies similar reasoning to discrete-time dynamics.

We first define a local figure of merit for the atypical features of the short-orbit automaton eigenstate $\ket{\varepsilon}$.  Let $\mathcal O$ be a local operator such that $\bra{\varepsilon}\mathcal O\ket{\varepsilon}$ differs from the infinite-temperature thermal expectation value $\langle\mathcal O\rangle_{\infty}=(\text{tr}_{\mathcal H}\, \mathcal O)/\mathcal D$, where $\mathcal D$ is the dimension of the Hilbert space $\mathcal H$.  Such an operator is guaranteed to exist for any $\ket{\varepsilon}$: since $\ket{\varepsilon}$ is a short-orbit automaton eigenstate and therefore has constant entanglement entropy, it can be expressed as a finite-bond-dimension matrix product state. Since such states are generically frustration-free eigenstates of a local Hamiltonian~\cite{Fannes92,Nachtergaele96,PerezGarcia07,Fernandez15}, one can always find a local operator $\mathcal O$ for which $|\bra{\varepsilon}\mathcal O\ket{\varepsilon}|$ is a number of order one, using, e.g., any of the local projectors entering the state's parent Hamiltonian.  As an example, for $\ket{\varepsilon}=\ket{\varepsilon_{3,p};\Omega}$, any of the vacuum-orbit eigenstates of the PXP automaton~\eqref{eq:Uaut}, the operator $\mathcal O=Z_iZ_{i+1}$ has $\bra{\varepsilon}\mathcal O\ket{\varepsilon}=-1/3$, while $\langle\mathcal O\rangle_{\infty}=2/\sqrt5-1\approx -0.11$ for the Fibonacci Hilbert space in the limit $L\to\infty$~\cite{Schecter18}.

We now wish to provide a bound on the quantity
\begin{align}
\begin{split}
     |\delta\mathcal O(t)|&=|\bra{\varepsilon}\mathcal O(t)\ket{\varepsilon}-\bra{\varepsilon}\mathcal O\ket{\varepsilon}|\\
    &=|\bra{\varepsilon}(U^\dagger)^t\mathcal O\, U^t\ket{\varepsilon}-\bra{\varepsilon}\mathcal O\ket{\varepsilon}|.
\end{split}
\end{align}
For a Floquet operator of the form $U=e^{-i\, \lambda\, H}\, U_{\rm aut}$ with $\lambda\ll 1$, we have
\begin{align}
\begin{split}
    \bra{\varepsilon}\mathcal O(t+1)\ket{\varepsilon} &= \bra{\varepsilon}U^\dagger\mathcal O(t)U\ket{\varepsilon}\\
    &\approx \bra{\varepsilon}\mathcal O(t)\ket{\varepsilon}+i\lambda\bra{\varepsilon}[H,\mathcal O(t)]\ket{\varepsilon},
\end{split}
\end{align}
to leading order in $\lambda$, where we used the fact that $\ket{\varepsilon}$ is an eigenstate of the unitary operator $U_{\rm aut}$. (Note that $\lambda \ll 1$ is the only approximation used in these manipulations.) Applying this formula recursively down to $t=0$ gives
\begin{align}
    \bra{\varepsilon}\mathcal O(t)\ket{\varepsilon}=\bra{\varepsilon}\mathcal O\ket{\varepsilon} + i\lambda\sum^{t-1}_{k=0}\bra{\varepsilon}[H,\mathcal O(k)]\ket{\varepsilon}, 
\end{align}
which leads to
\begin{align}
\label{eq:OSumBound}
\begin{split}
    |\delta \mathcal O(t)|
&\approx \lambda\, \Big|\sum^{t-1}_{k=0}\bra{\varepsilon}[H,\mathcal O(k)]\ket{\varepsilon}\Big|\\
&\leq\lambda\, \sum^{t-1}_{k=0}|\bra{\varepsilon}[H,\mathcal O(k)]\ket{\varepsilon}|.
\end{split}
\end{align}
We can now use the fact that $\mathcal O(k)$ is a local operator evolved by the finite-depth local quantum circuit $U^k$, and hence the support of $\mathcal O(k)$ must lie within a ``light cone" whose size scales as $k$.  Since $H$ is a local Hamiltonian, this implies the bound
\begin{align}
\label{eq:LightCone}
    |\bra{\varepsilon}[H,\mathcal O(k)]\ket{\varepsilon}|\leq c_{0,k}+c_{1,k}\, k,
\end{align}
where $c_{0,k}$ and $c_{1,k}$ are order-one constants. Then we can bound the sum in Eq.~\eqref{eq:OSumBound} as
\begin{align}
\label{eq:Sum}
\begin{split}
    \sum^{t-1}_{k=0}|\bra{\varepsilon}[H,\mathcal O(k)]\ket{\varepsilon}|
   &\leq \sum^{t-1}_{k=0}\left(c_{0,k}+c_{1,k}\, k\right)\\
   &\leq c_0\, t+c_1\sum^{t-1}_{k=0} k\\
   &= c_0\, t+c_1\, \frac{t(t-1)}{2},
\end{split}
\end{align}
where $c_0=\text{max}_k\, c_{0,k}$ and $c_1=\text{max}_k\, c_{1,k}$.  We thus arrive at the bound
\begin{align}
    |\delta \mathcal O(t)|\leq \lambda\, (c'_0\, t+ c'_1\, t^2),
\end{align}
where $c'_0=c_0-c_1/2$ and $c'_1=c_1/2$. The above bound implies that $|\delta \mathcal O(t)|$ does not become of order one before a time $t_*\sim \lambda^{-1/2}$.  Thus, the time scale for the relaxation of $\mathcal O$ to its infinite-temperature value is bounded from below by $t_*$, which becomes arbitrarily large as $\lambda$ is made arbitrarily small.  

The above result is in agreement with that of Ref.~\cite{Lin19}, which applies to continuous-time Hamiltonian dynamics. Our result can be generalized to $d$ spatial dimensions along the lines of Ref.~\cite{Lin19}, which finds that $t_*\sim \lambda^{-1/(d+1)}$.  This is because the volume of the light cone in Eq.~\eqref{eq:LightCone} scales like $k^d$ in $d$ dimensions.  Then $\sum^{t-1}_{k=0} k^d$ in the generalization of Eq.~\eqref{eq:Sum} can be evaluated using Faulhaber's formula to obtain a degree-$(d+1)$ polynomial in $t$.

Finally, we note that the lower bound on the thermalization time scale presented here is likely quite loose, as it does not take into account any structure in the perturbation $H$ beyond spatial locality.  For example, the thermalization time scale evident in Fig.~\ref{fig:dynamics}(a) in the main text is much larger than the prediction $t_*\sim \lambda^{-1/2}$.

\section{Derivation of the PXP Circuit from the Time-Dependent Ising Model}
\label{sec:Ising_Floquet_circuit}

In this Appendix we show how to derive Eq.~\eqref{eq:UF} from the time-dependent long-range Ising model \eqref{eq:H}. We consider the Floquet dynamics generated by
\begin{subequations}
\begin{align}\label{eq:UF-ising}
U^{\rm Ising}_{\rm F}=e^{i\, \frac{T}{2}\, H_{\rm odd}}e^{i\, \frac{T}{2}\, H_{\rm even}},
\end{align}
where
\begin{align}
H_{\rm even(odd)}
&= 
\sum_{i,j}V_j\, n_{i}n_{i+j}
+h_0
\!\!\!\sum_{i\text{ even(odd)}}X_i.
\end{align}
\end{subequations}
Physically, this circuit corresponds to a series of repeated quenches of the Hamiltonian \eqref{eq:H} in which the transverse field is applied to sites on the even and odd sublattices in an alternating fashion.  The corresponding time-dependent Hamiltonian is given by
\begin{subequations}
\begin{align}
H(t)
&= 
\sum_{i,j}V_j\, n_{i}n_{i+j}
+h_0\left[
f(t)\sum_{i\text{ even}}X_i
+
g(t)\sum_{i\text{ odd}}X_i
\right],
\end{align}
where
\begin{align}
f(t) = \frac{\text{sgn}(\sin\omega t)+1}{2},\indent g(t)=1-f(t)
\end{align}
are square pulses and $\omega=2\pi/T$; this is precisely Eq.~\eqref{eq:H} with the square wave drive~\eqref{eq:SquareWave}.
\end{subequations}
We claim that, in the limit $V_1 \gg h_0$ and $V_j=0$ for $j>1$, the dynamics under this circuit matches that of Eq.~\eqref{eq:UF}.  Moreover, as shown in the main text, the qualitative features of this dynamics survives the experimentally realistic inclusion of longer-range interactions, such as $V_2 = V_1/2^6$.

To see how Eq.~\eqref{eq:UF} emerges from Eq.~\eqref{eq:UF-ising}, we first write
\begin{subequations}
\begin{align}
H(t) = V_1\left\{H_0+\left(\frac{h_0}{V_1}\right)\left[W(t)+W^\prime(t)\right]\right\},
\end{align}
where
\begin{align}
H_{0}
&= 
\sum_{i,j}\frac{V_j}{V_1}\, n_{i}n_{i+j},\\
W(t)&=\mathcal P_{\rm fib}\left[
f(t)\sum_{i\text{ even}}X_i
+
g(t)\sum_{i\text{ odd}}X_i
\right]
\mathcal P_{\rm fib},
\\
W^\prime(t)&=f(t)\sum_{i\text{ even}}X_i
+
g(t)\sum_{i\text{ odd}}X_i
-W(t),
\end{align}
with $\mathcal P_{\rm fib}=\frac{1}{2}\prod_i(1-n_in_{i+1})$ the projector onto the Fibonacci Hilbert space.  This decomposition is such that $W(t)$ is block-diagonal in the Fibonacci Hilbert space and $W^\prime(t)$ is block-off-diagonal.
\end{subequations}
Next, we perform degenerate perturbation theory within the Fibonacci Hilbert space based on a \textit{time-dependent} Schrieffer-Wolff transformation generated by an anti-Hermitian operator $S(t)$:
\begin{widetext}
\begin{subequations}
\begin{align}
H_{\rm eff}(t)&=e^{S(t)}[H(t)-i\partial_t]e^{-S(t)}\\
&=V_1H_0+h_0\left[W(t)+W^\prime(t)\right]+V_1[S(t),H_0]+h[S(t),W(t)]+h_0[S(t),W^\prime(t)]+\frac{1}{2}V_1[S(t),[S(t),H_0]]+\dots\\
&\qquad\qquad+i \dot{S}(t)+i[S(t),\dot{S}(t)]+i\frac{1}{2}[S(t),[S(t),\dot{S}(t)]+\dots,
\end{align}
\end{subequations}
\end{widetext}
where $\dot{S}(t)\equiv\partial_tS(t)$ and where the last line originates from the contribution $e^{S(t)}(-i\partial_t)e^{-S(t)}$, which arises due to the time-dependence of $S(t)$.  Next, we expand $S(t)$ in powers of $h_0/V_1$ as
\begin{widetext}
\begin{align}
S(t)=\sum^{\infty}_{n=1}\left(\frac{h_0}{V_1}\right)^n\,S_n(t)
\end{align}
and collect terms by their order in $h_0/V_1$:
\begin{subequations}
\begin{align}
H_{\rm eff}(t)&=
V_1H_0+h_0\left\{W(t)+W^\prime(t)+[S_1(t),H_0]+i\frac{1}{V_1}\dot{S}_1(t)\right\}\\
&\qquad+\frac{h^2_0}{V_1}\left\{[S_2(t),H_0]+[S_1(t),W(t)]+[S_1(t),W^\prime(t)]+\frac{1}{2}[S_1(t),[S_1(t),H_0]]+i\frac{1}{V_1}\dot{S}_2(t)+i\frac{1}{V_1}[S_1(t),\dot{S}_1(t)]\right\}+\dots\\
&\equiv
V_1\sum^\infty_{n=0}\left(\frac{h_0}{V_1}\right)^nH_n.
\end{align}
\end{subequations}
\end{widetext}
The task is to choose $S_n(t)$ at each order such that $[H_n,\mathcal P_{\rm fib}]=0$
(note that this is automatically true of $H_0$).  To do this at leading order in $h$, we require
\begin{align}
W^\prime(t)+[S_1(t),H_0]+i\frac{1}{V_1}\dot{S}_1(t)=0.
\end{align}
To solve this equation, we recall that $W^\prime(t)$ is a periodic function of $t$ with period $T=2\pi/\omega$.  This motivates the introduction of the Fourier decomposition of an operator $\mathcal O(t)$,
\begin{subequations}
\begin{align}
\mathcal O(t)&=\sum^{\infty}_{m=-\infty}\mathcal O^{(m)}\, e^{-im\omega t}\\
\mathcal O^{(m)}&=\frac{1}{T}\int^T_0 dt\, e^{im\omega t}\mathcal O(t),
\end{align}
\end{subequations}
which we substitute into the above to obtain the equations
\begin{align}
W^{\prime\, (m)}+[S^{(m)}_1,H_0]+\frac{m\omega}{V_1}S^{(m)}_1=0
\end{align}
for each $m$.  Next we write the above equation in the eigenbasis of $H_0$,
$V_1 H_0\ket{\sigma}=E_{\sigma}\ket{\sigma}$,
and solve for the matrix elements of $S^{(m)}_1$:
\begin{align}
\bra{\sigma}S^{(m)}_{1}\ket{\sigma'}=V_1\frac{\bra{\sigma}W^{\prime\, (m)}\ket{\sigma'}}{E_\sigma-E_{\sigma'}-m\omega}.
\end{align}
Note that we must require $E_\sigma-E_{\sigma'}\neq m\omega$ for any $m$ in order for this solution to be well defined; in practice this requirement is not very restrictive.

We have thus shown that, to leading order in an expansion in $h_0/V_1$,
\begin{widetext}
\begin{align}
\label{eq:NNBlockadeH}
H_{\rm eff}(t)\approx V_1 H_0+h_0\left[f(t)\sum_{i\text{ even}}(PXP)_i
+
g(t)\sum_{i\text{ odd}}(PXP)_i\right]+\dots,
\end{align}
which corresponds to a Floquet operator
\begin{align}
U_{\rm F}\approx e^{-i\frac{h_0T}{2}\left[\frac{V_1}{h_0}H_0+\sum_{j\, \mathrm{odd}}(PXP)_j\right]}\, e^{-i\frac{h_0T}{2}\left[\frac{V_1}{h_0}H_0+\sum_{j\, \mathrm{even}}(PXP)_j\right]}.
\end{align}
\end{widetext}
Thus we see that, if $H_0$ consisted solely of nearest-neighbor interactions (i.e.~$V_{n\geq 2}=0$), the Floquet unitary $U_{\rm F}$ corresponding to $H_{\rm eff}(t)$ would be precisely Eq.~\eqref{eq:UF} with $\theta=h_0T/2$. (Note that in this case $H_0$ vanishes when acting on any state in the Fibonacci subspace.)  In the realistic scenario where $V_{n}\sim V_1/n^6$, $U_{\rm F}$ contains corrections to the underlying automaton dynamics even at leading order in $h_0/V_1$ due to the further-neighbor couplings.  However, our numerical results indicate that these corrections do not preclude the ability to observe features of the underlying automaton dynamics at $\theta=\pi/2$.

\section{Family of CAs from Further-Neighbor Blockade}
\label{sec:FurtherNeighbor}
Appendix~\ref{sec:Ising_Floquet_circuit} demonstrates how the PXP circuit \eqref{eq:UF} emerges from the driven long-range Ising model \eqref{eq:H} in the limit $V_1\gg h_0 \gg V_{r\geq 2}$.  The reduction of the long-range Ising model \eqref{eq:H} to a model of the form \eqref{eq:NNBlockadeH} that commutes with the projector $\mathcal P_{\rm fib}$ is referred to in the AMO literature as ``Rydberg blockade,"~\cite{Jaksch00}.  In this regime, excited Rydberg atoms (``up" spins) carry an excluded volume of radius $r=1$ site due to strong nearest-neighbor interactions.  However, it is well known that the ``blockade radius" $r$ can be increased by allowing large further-neighbor couplings.  In this Appendix we show how this mechanism can be used to generate a family of automaton dynamics that generalizes the PXP automaton discussed in the main text.

The $r$th-neighbor blockade regime is defined by the hierarchy of scales $V_1>\ldots>V_r\gg h_0\gg V_{r+1}>\dots,$ which is ultimately controlled by $V_1$ in the case of van der Waals interactions~\cite{Bernien17}.  This assigns a large energetic penalty not just to nearest-neighbor pairs of excited Rydberg atoms, but to further-neighbor pairs as well.  This induces a blockaded subspace of the full Hilbert space that generalizes the Fibonacci subspace and is defined by the projector
\begin{align}
    \mathcal P_{r}=\frac{1}{2^r}\prod^r_{q=1}\prod_j (1-n_in_{i+q})
\end{align}
(note that $\mathcal P_1=\mathcal P_{\rm fib}$).  The quantum dynamics generated by the time-dependent long-range Ising model \eqref{eq:H} can be projected into this blockaded subspace by a direct generalization of the calculation in Appendix~\ref{sec:Ising_Floquet_circuit} in which $\mathcal P_{\rm fib}$ is replaced by $\mathcal P_r$.  The result is a generalization of Eq.~\eqref{eq:UF} in which the operator $(PXP)_j$ is replaced by
\begin{align}
    (P^rXP^r)_j=\left(\prod^{r}_{q=1} P_{j-q}\right)X_j\left(\prod^{r}_{s=1} P_{j+s}\right).
\end{align}
Tuning the angle $\theta=h_0T/2$ to $\pm \pi/2$ mod $2\pi$ then yields a classical automaton in which the gate $\mathsf{Toffoli}_j(\phi)$ is replaced by the gate
\begin{align}
    \mathsf{C}^{2r}\mathsf{NOT}_j(\phi)=1-\left(\prod^{r}_{q=1} P_{j-q}P_{j+q}\right)+e^{i\phi}(P^rXP^r)_j.
\end{align}
In other words, as the blockade radius $r$ increases, the number of control qubits in the classical update rule grows.

\section{Scaling of the Vacuum Orbit Weight and Entanglement Entropy}

\begin{figure}[t!]
\begin{center}
\includegraphics[width=.85\columnwidth]{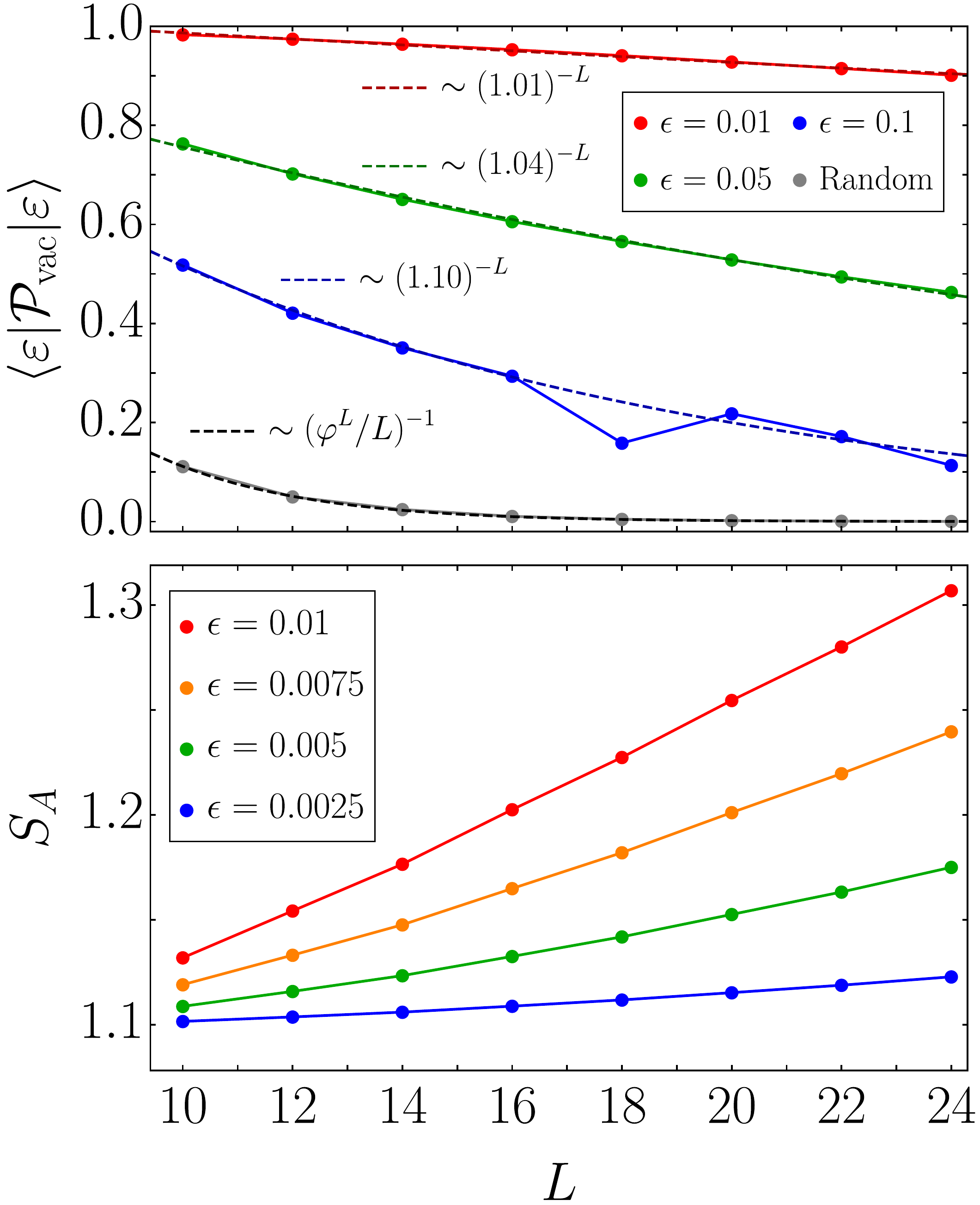}\\
\caption{
Finite-size and -$\epsilon$ scaling of the vacuum orbit weight (top) and the half-chain entanglement entropy (bottom) for the entanglement outlier closest to quasienergy $\varepsilon=-2\pi/3$ in the PXP automaton with PBC.
}
\label{fig:pxp-eigenstate}
\end{center}
\end{figure}

The low-entanglement eigenstates of the perturbed PXP circuit visible in Fig.~\ref{fig:pxp-entanglement} in the main text are precisely those with the highest weight on the orbit of the vacuum state $\ket{\Omega}$, quantified by the projection operator $\mathcal P_{\rm vac}=\sum^2_{k=0}U^k_{\rm aut}\ket{\Omega}\bra{\Omega}(U^\dagger_{\rm aut})^k$. The expectation value of $\mathcal P_{\rm vac}$ for the entanglement outlier closest to $\varepsilon=-2\pi/3$ is plotted as a function of $\epsilon$ and $L$ in the top panel of Fig.~\ref{fig:pxp-eigenstate}.  Points on the red curve for $L=20,22,24$ correspond to the highlighted entanglement outliers in Fig.~\ref{fig:pxp-entanglement}(a) in the main text, demonstrating their anomalously large weight ($\gtrsim0.9$) on the vacuum orbit. For comparison, a random state in the zero-momentum sector has exponentially small weight $\sim 3/(\varphi^L/L)$ (where $\varphi\approx1.618$ is the golden ratio) on the same orbit (gray curve). At small $\epsilon$, the vacuum orbit weight fits well to an exponential form $\sim(1+\epsilon)^{-L}$, corresponding to an exponential enhancement compared to the random case.  As $\epsilon$ increases, a crossover to random-state scaling occurs that is likely driven by many-body resonances.  The effect of such a resonance is seen on the $\epsilon=0.1$ curve, where an outlier appears at $L=18$.  We have verified that there is another eigenstate very close in quasienergy to this outlier with almost the same orbit weight, suggesting that a resonance is the underlying cause.  Such resonances also appear in studies of QMBS in the PXP model~\cite{Turner18b,Iadecola19}, and are believed to underlie their disappearance in the thermodynamic limit~\cite{Lin19}.

The bottom panel of Fig.~\ref{fig:pxp-eigenstate} plots the half-chain entanglement entropy of the same low-entanglement eigenstates as a function of $L$ and $\epsilon$.  $S_A$ exhibits clear \emph{volume-law} scaling with $L$, albeit with a small slope of order $\epsilon$ for sufficiently small $\epsilon$.  This is consistent with the data in the top panel of Fig.~\ref{fig:pxp-eigenstate}: the eigenstate in question has a concentration of weight in the vacuum-orbit Hilbert space, and the part of its weight residing outside this subspace is essentially random.  Since only the part of the state in the complement of the vacuum subspace can contribute to volume-law entanglement, the volume-law slope is limited by the effective dimension of the resulting Hilbert space, which grows like $(1+\epsilon)^L$ for small $\epsilon$, as suggested by the top panel of Fig.~\ref{fig:pxp-eigenstate}. This leads to a rough estimate $S_A\sim L\ln(1+\epsilon)\approx \epsilon L$, consistent with the numerical data.

The above finite-size and small-$\epsilon$ scaling analysis suggests that these eigenstates ultimately become indistinguishable to numerical precision from typical eigenstates at sufficiently large system sizes, as is likely the case for the approximately scarred eigenstates of the PXP model~\cite{Choi18,Lin19}.  However, the deviations of these eigenstates' properties from ETH predictions are substantial even for relatively large but finite system sizes---for example, if the scaling predictions of Fig.~\ref{fig:pxp-eigenstate} hold for $\epsilon=0.01$, $L\sim250$ is required to drive the eigenstate weight on the vacuum sector below 10\%, which is still many orders of magnitude larger than the prediction for a random state.

\end{document}